\documentclass[dvips]{plain}
\usepackage{supertabular,lscape,epsfig}
\usepackage{amssymb}
\usepackage{amsmath}
\usepackage{wasysym}

\SetPages{0}{0}

\SetVol{0}{2020}

\begin{document}

\begin{Titlepage}
\Title{Planet-star interactions with precise transit timing. II. The radial-velocity tides and a tighter constraint on the orbital decay rate in the WASP-18 system}
\Author{G.~~M~a~c~i~e~j~e~w~s~k~i$^{1}$, ~ H.~~A.~~K~n~u~t~s~o~n$^{2}$, A.~~W.~~H~o~w~a~r~d$^{3}$, ~ H.~~I~s~a~a~c~s~o~n$^{4}$, ~ E.~~F~e~r~n~\'a~n~d~e~z~-~L~a~j~\'u~s$^{5,6}$, ~ R.~~P.~~D~i~~S~i~s~t~o$^{5,6}$, ~ C.~~M~i~g~a~s~z~e~w~s~k~i$^{1}$}
{$^1$Institute of Astronomy, Faculty of Physics, Astronomy and Informatics,
         Nicolaus Copernicus University, Grudziadzka 5, 87-100 Toru\'n, Poland,
         e-mail: gmac@umk.pl\\
 $^2$Division of Geological and Planetary Sciences, California Institute of Technology, 1200 E. California Blvd., Pasadena, CA 91125, USA\\
 $^3$Institute for Astronomy, University of Hawaii, Honolulu, HI 96822, USA\\
 $^4$Department of Astronomy, University of California Berkeley, Berkeley CA 94720, USA\\
 $^5$Facultad de Ciencias Astron\'omicas y Geof\'isicas, Universidad Nacional de La Plata, 1900 La Plata, Buenos Aires, Argentina\\
 $^6$Instituto de Astrof\'isica de La Plata (CCT La Plata C CONICET/UNLP), Paseo del Bosque s/n, La Plata, BA, B1900FWA, Argentina}

\Received{April, 2020}
\end{Titlepage}

\Abstract{From its discovery, the WASP-18 system with its massive transiting planet on a tight orbit was identified as a unique laboratory for studies on tidal planet-star interactions. In an analysis of Doppler data, which include five new measurements obtained with the HIRES/Keck-I instrument between 2012 and 2018, we show that the radial velocity signal of the photosphere following the planetary tidal potential can be distilled for the host star. Its amplitude is in agreement with both theoretical predictions of the equilibrium tide approximation and an ellipsoidal modulation observed in an orbital phase curve. Assuming a circular orbit, we refine system parameters using photometric time series from TESS. With a new ground-based photometric observation, we extend the span of transit timing observations to 28 years in order to probe the rate of the orbital period shortening. Since we found no departure from a constant-period model, we conclude that the modified tidal quality parameter of the host star must be greater than $3.9 \times 10^6$ with 95\% confidence. This result is in line with conclusions drawn from studies of the population of hot Jupiters, predicting that the efficiency of tidal dissipation is 1 or 2 orders of magnitude weaker. As the WASP-18 system is one of the prime candidates for detection of orbital decay, further timing observations are expected to push the boundaries of our knowledge on stellar interiors.}{planet-star interactions -- stars: individual: WASP-18 -- planets and satellites: individual: WASP-18 b}

%%%%%%%%%%%%%%%%%%%%%%%%%%%%%%%%%%%%%%%%%%%%%%%%%%%%%%%%%%%%%%%%%%%%%%

\section{Introduction}

Hot Jupiters, i.e. massive exoplanets on extremely tight orbits, are recognised as unique laboratories for studying planet--star interactions. Their orbital distances of 0.02--0.03 AU are small enough to produce detectable tidal deformations of their host stars. Departure from a spherical symmetry of the gravitational potential gives a rise to apsidal precession that could be detected via transit and occultation timing (\eg Ragozzine \& Wolf 2009) and radial velocity (RV) variations (\eg Csizmadia \etal 2019). Dissipation of energy deposited in stellar tides leads to orbital shrinkage (e.g.\ Levrard \etal 2009), which is also observationally accessible via long-term transit and occultation timing. 

Tidal deformation of a host star can be observed with both the photometric and Doppler techniques. Tidal ellipsoidal distortions follow the orbital motion of a planet and modulate both an observed light curve and RV signal with a period being half the orbital period. In the most favourable circumstances, amplitudes of these modulations are of order $10^2$ parts per milion (ppm) of the observed flux (Pfahl \etal 2008) and a few m~s$^{-1}$ in the Doppler domain (Arras \etal 2012). 

The WASP-18 system comprises an F/G dwarf ($V=9.3$ mag) being orbited by a 10-$M_{\rm{Jup}}$ planet on a 0.94-day orbit (Hellier \etal 2009). Recent astrometric studies suggest that there is a $\sim$0.1 $M_{\odot}$ companion at a projected separation of $\sim$3300 AU (Csizmadia \etal 2019, Fontanive \etal 2019). The host star is referred to as WASP-18A, while the secondary component of that binary system is named WASP-18B. WASP-18A~b was immediately recognised as a promising candidate for an in-falling planet. Using a canonical value of the modified tidal quality parameter of the host star $Q'_{\star} = 10^6$ (Meibom \& Mathieu 2005, Ogilvie \& Lin 2007, Milliman \etal 2014), which characterises the efficiency of dissipation of tidal energy, the remaining system lifetime would be of about $7\times10^5$~yr and a cumulative departure from a linear transit ephemeris would reach  about 30~s after 10 years (Hellier \etal 2009). The WASP-18 system was also identified as a prime candidate for which an RV signature of planetary induced tidal deformations could be detected (Arras \etal 2012). The orbit of WASP-18A~b appears to be slightly eccentric with a line of apsides orientated along the line of sight. Such a specific configuration actually corresponds to a tidal RV signal produced by a fluid flow in the star forced by the massive planetary companion. Those ellipsoidal distortions of WASP-18A were found to produce photometric modulation in a visible-light orbital phase curve with an amplitude of about 200 ppm (Shporer \etal 2019).

In this paper, we show that the tidal RV signal can be distilled from the observed RV variations for WASP-18A, and its amplitude is consistent with theoretical expectations. We also use the transit timing data, spanning 28 years, to search for the orbital shrinkage for WASP-18A~b and to constrain the efficiency of tidal dissipation in its host star. 
 
%%%%%%%%%%%%%%%%%%%%%%%%%%%%%%%%%%%%%%%%%%%%%%%%%%%%%%%%%%%%%%%%%%%%%%

\section{RV reanalysis}

We acquired 5 RV measurements with the High Resolution Echelle Spectrometer (HIRES, Vogt \etal 1994) mounted at the 10 m Keck I telescope as an extension of the observing programme presented in Knutson \etal (2014).  
The original dataset comprises 6 RV observations secured between 2010 February and October 2012. Their errors were in a range 3.7--5.6 m~s$^{-1}$ with a median value of 4.0 m~s$^{-1}$. Our new observations were performed between October 2012 and August 2019. The instrumental setup and data-reduction pipeline was adopted from the California Planet Search consortium (Wright \etal 2004, Howard \etal 2009, Johnson \etal 2010, see also Knutson \etal 2014 for details). For the sake of homogeneity, the previously published HIRES RVs were also reprocessed. Mid-exposures were converted to barycentric Julian dates in barycentric dynamical time $(\rm{BJD_{TDB}})$. The complete dataset is given in Table~1.  

Additional RV measurements were taken from Triaud \etal (2010) and Albrecht \etal (2012). Thirty seven of them were acquired between 2007 September and 2009 January using the CORALIE high resolution \'echelle spectrograph paired with the 1.2 m Euler Swiss Telescope (La Silla, Chile). The errors were in a range 8.2--14.2 m~s$^{-1}$ with a median value of 9.7 m~s$^{-1}$. Forty eight measurements come from the Planet Finder Spectrograph (PFS) and the Magellan II 6.5 m telescope (Las Campanas, Chile). They were gathered on a single night in October 2011 in order to study the Rossiter--McLaughlin (RM) effect. They have errors between 4.9 and 9.2 m~s$^{-1}$ with a median value of 5.9~m~s$^{-1}$. Twenty three measurements were secured in 2008 August with the high resolution \'echelle spectrograph HARPS at the 3.6 m ESO telescope at La Silla. They were also used to investigate the RM effect. The errors were between 4.4 and 10.7 m~s$^{-1}$ with a median value of 6.1~m~s$^{-1}$.

\MakeTable{ l c c c }{12.5cm}{New and reprocessed RV measurements acquired with HIRES/Keck I.}
{\hline
 Mid-exposure $(\rm{BJD_{TDB}})$ & Relative RV$^{*}$ m~s$^{-1}$ & RV error$^{*}$ (m~s$^{-1}$) & Remarks \\
  & & & \\
\hline
2455231.725706 & $-311.832$  & 5.828 & (1) \\
2455427.048972 & $909.848$   & 3.519 & (1) \\
2456167.071476 & $1371.783$  & 3.951 & (1) \\
2456193.094044 & $-1341.056$ & 4.255 & (1) \\
2456197.032589 & $-553.433$  & 4.344 & (1) \\
2456207.967066 & $129.533$   & 4.526 & (2) \\
2456209.021271 & $-946.126$  & 4.685 & (1) \\
2456913.036796 & $1092.56$4  & 4.290 & (2) \\
2457241.125035 & $-335.939$  & 4.793 & (2) \\
2458393.977843 & $797.761$   & 2.886 & (2) \\
2458720.096038 & $-815.098$  & 5.313 & (2) \\
\hline
\multicolumn{4}{l}{Remarks: (1) reprocessed from Knutson \etal (2014), (2) new measurement} \\
\multicolumn{4}{l}{$^{*}$ higher numerical precision left intentionally.} \\
}

The procedure, we followed, was adopted from Maciejewski \etal (2020). Thirty seven RV data points fall in transit phase and were affected by the RM effect. Since our analysis procedure does not take this effect into account, the appropriate corrections were calculated using the RM model obtained by Albrecht \etal (2012) and then subtracted from the original RV measurements. The final RV sample comprised 119 RV data points. In order to place a constraint on a transit ephemeris, a set of mid-transit times (Sect.~4) was added together with mid-occultation times taken from Nymeyer \etal (2011), Maxted \etal (2013), and Wilkins \etal (2017), corrected for the light-travel time across the WASP-18A~b's orbit. The Keplerian model of the orbit was characterised by 9 free parameters: the orbital period $P_{\rm{orb}}$, RV amplitude $K_{\rm{orb}}$, mean anomaly for a given epoch, apparent orbital eccentricity $e_{\rm{orb}}$, longitude of periastron $\omega$, and four RV offsets for the individual RV datasets. The best-fitting Keplerian solution was found with the Levenberg-Marquardt algorithm. The parameters' uncertainties were estimated with the bootstrap method using the median absolute deviations for $10^6$ resampled datasets.

As compared to a circular-orbit model, the eccentric model is favoured by the Bayesian information criterion (BIC),
\begin{equation}
  {\rm{BIC}} = {\chi}^2 + k \ln N,
\end{equation}
where $k$ is the number of fit parameters and $N$ is the number of data points, with $\Delta \rm{BIC} \approx 330$. The RV residuals against the best-fitting solution exhibit unmodeled data-point scatter at the level of 10.7~m~s$^{-1}$. This RV jitter was added in quadrature to the RV errors in the final iteration. We found $e_{\rm{orb}} = 0.010 \pm 0.001$ and $\omega = 268.7^{\circ} \pm 1.4^{\circ}$. 

Arras \etal(2012) demonstrated that an orbital configuration with a nonzero eccentricity and longitude of periastron close to $270^{\circ}$ might be de facto a signal comprising a circular orbit component $V_{\rm{orb}}(\phi)$ and a tidal component $V_{\rm{tide}}(\phi)$, where $\phi$ is an orbital phase. The observed RV signal $V_{\rm{obs}}(\phi)$ can be written as 
\begin{equation}
  V_{\rm{obs}}(\phi) = \gamma + V_{\rm{orb}}(\phi) + V_{\rm{tide}}(\phi)\, , \;
\end{equation}
where $\gamma$ is the barycentre velocity and 
\begin{equation}
  V_{\rm{orb}}(\phi) = - K_{\rm{orb}} \sin \left( 2 \pi (\phi - \phi_{0}) \right) \, , \;
\end{equation}
and 
\begin{equation}
  V_{\rm{tide}}(\phi) = K_{\rm{tide}} \sin \left( 4 \pi (\phi - \phi_{0}) \right)\, . \;
\end{equation}
In those formulae, $\phi_{0}$ is the phase offset for a reference epoch, and $K_{\rm{orb}}$ and $K_{\rm{tide}}$ are the amplitudes of the orbital motion and the tidal component, respectively. The best-fitting solution was found with the Markov chain Monte Carlo (MCMC) employing the \textit{emcee} sampler (Foreman-Mackey \etal 2013). One hundred walkers, $10^5$ steps long each, were used to produce marginalised posteriori probability distributions for the free parameters. The first 10\% of steps were discarded in a burn-in phase. Median values of the cumulative distributions were taken as the best fitting parameters, and 15.9 and 84.1 percentile values of those distributions were used as the 1$\sigma$ uncertainties. 

The best fitting model is presented in panel (a) in Fig.~1 together with the residuals in panel (b). The orbital RV component is plotted in panel (c), and the tidal RV signal is shown in panel (d). We obtained $K_{\rm{orb}}=1813.9\pm2.4$~m~s$^{-1}$ and $K_{\rm{tide}}=17.9\pm1.9$~m~s$^{-1}$. The phase offsets $\phi_{0}=(4 \pm 15) \times 10^{-5}$ was found to be consistent with zero well within 1$\sigma$. Although the barycentre velocity was subtracted from the RV measurements prior to the fitting procedure, its uncertainty was taken into account in the error budget by allowing an RV shift to float. This shift was found to be $-0.3\pm1.4$~m~s$^{-1}$, \ie consistent with zero well within 1$\sigma$.
 
% FIGURE 
\begin{figure}[thb]
\begin{center}
\includegraphics[width=1.0\textwidth]{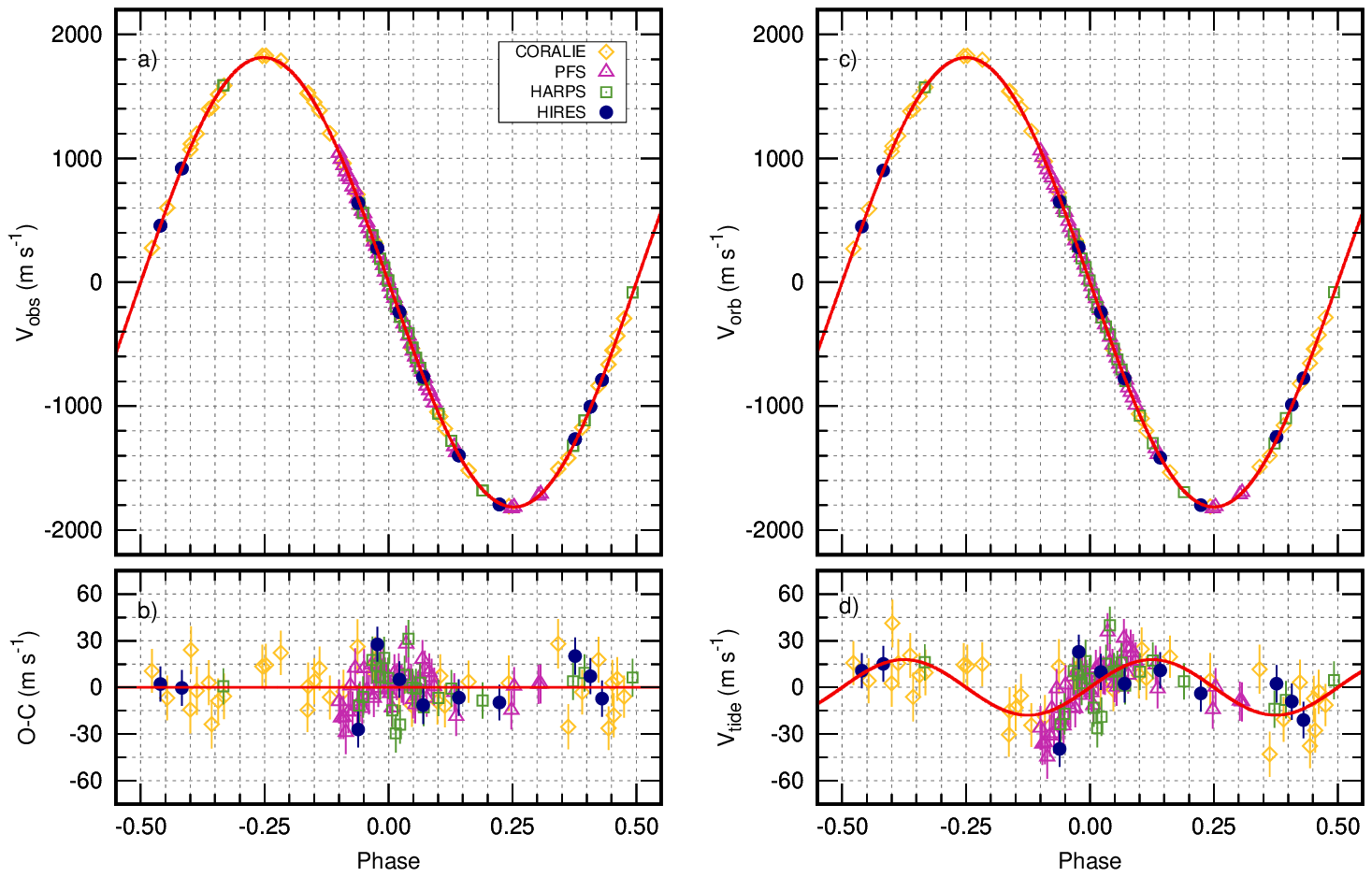}
\end{center}
\FigCap{\textit{Panel (a)}: relative RV measurements observed for WASP-18A, phase-folded with the orbital period of WASP-18A~b. Open symbols mark measurements taken from the literature. Dots show our new and reanalysed observations acquired with HIRES. The error bars of individual measurements are increased by the value of jitter of 10.7 m~s$^{-1}$, added in quadrature. The red line shows the best-fitting model comprises two components: the orbital motion of the planet on a circular orbit and the tidal RV signal. \textit{Panel (b)}: the residuals from the best-fitting model. \textit{Panel (c)}: orbital RV component. \textit{Panel (d)}: tidal RV component.}
\end{figure}

%%%%%%%%%%%%%%%%%%%%%%%%%%%%%%%%%%%%%%%%%%%%%%%%%%%%%%%%%%%%%%%%%%%%%%

\section{TESS photometry}

The space-borne photometry from the Transiting Exoplanet Survey Satellite (TESS, Ricker \etal 2014) was used by Shporer \etal (2019) to refine system orbital parameters under the assumption that the orbit of WASP-18A~b is noncircular. As the observed eccentricity is likely a manifestation of the tidal RV signal, we reanalysed the photometric time series from TESS to refine system parameters for a circular orbit scenario.

TESS observed the WASP-18 system in sectors 2 (from August 22 to September 20, 2018) and 3 (from September 20 to October 18, 2018) with Camera 2. The photometric data of a 2 minute cadence were downloaded through the exo.MAST portal\footnote{https://exo.mast.stsci.edu}. The Presearch Data Conditioning (PDC) light curve was extracted for further analysis. A median value of recorded counts was calculated for each sector separately and then used for light-curve normalisation. To remove variability other than transits, a 12 hour boxcar was applied with in-transit and in-occultation data points masked. A transit ephemeris from Shporer \etal (2019) was used to extract data collected in transit windows and extended by 90 minutes of out-of-transit observations before and after each event. The set of 47 complete transit light curves was prepared for modelling with the Transit Analysis Package (TAP, Gazak \etal 2012).

Since the photometric time series might still be affected by out-of-transit variations, the TAP code was modified to be capable to model flux trends in a time domain with a second-order polynomial. In our approach, trends, as well as mid-transit times $T_{\rm{mid}}$, were modelled separately for each transit light curve. Transit parameters, such as the orbital inclination $i_{\rm{orb}}$, the semi-major axis scaled in star radii $a/R_{\star}$, and the ratio of planet to star radii $R_{\rm{p}}/R_{\star}$, were linked together for all light curves. The value of $P_{\rm{orb}}$ was taken from the transit-timing analysis (Sect.~4). The coefficients of the quadratic limb-darkening (LD) law -- the linear $u_{\rm{lin}}$ and the quadratic $u_{\rm{quad}}$ -- were allowed to float. Their initial values were bi-linearly interpolated from tables of Claret \& Bloemen (2011).

The best-fitting parameters and their uncertainties were determined from the marginalised posteriori probability distributions produced from 10 MCMC chains (\ie the median value, and 15.9 and 84.1 percentiles). The random walk process was driven by the Metropolis-Hastings algorithm and a Gibbs sampler. The wavelet-based technique (Carter \& Winn 2009) was employed to account for the correlated noise. Each chain was $10^6$ steps long. The first 10\% of trials were rejected to minimise the influence of the initial values. The best-fitting model is plotted in Fig.~2, and its parameters are listed in Table~2. We also give the results reported by Shporer \etal (2019) for comparison purposes. 

% FIGURE 
\begin{figure}[thb]
\begin{center}
\includegraphics[width=0.7\textwidth]{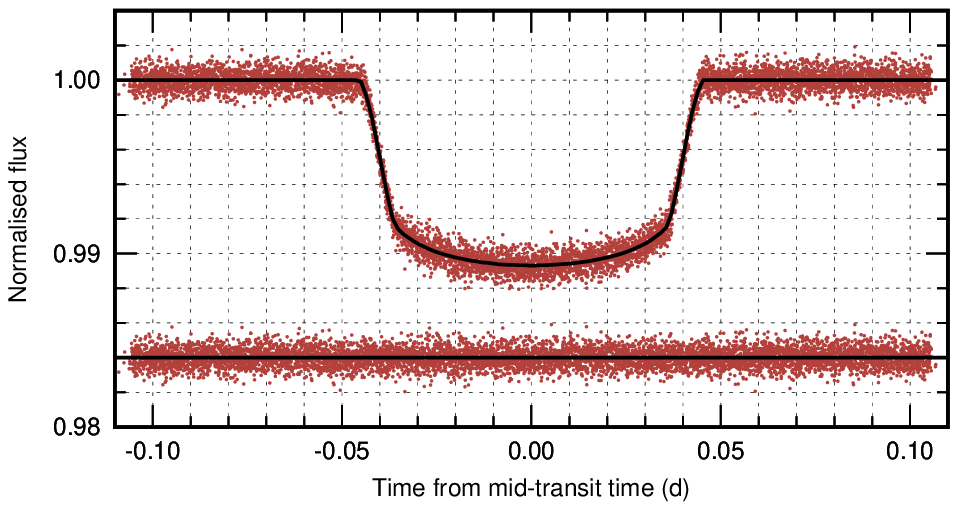}
\end{center}
\FigCap{Phase folded transit light curve from TESS with the best-fitting model. The residuals are plotted below.}
\end{figure}

\MakeTable{ l c c c }{12.5cm}{Transit parameters for WASP-18A~b re-determined for the circular orbit. The results from Shporer \etal (2019) and from the trial noncircular scenario are given for comparison purposes.}
{\hline
 Parameter & Circular & Shporer \etal (2019) & Noncircular \\
\hline
$R_p/R_{\star}$ & $0.09776^{+0.00028}_{-0.00027}$ & $0.09716^{+0.00014}_{-0.00013}$ & $0.09721^{+0.00018}_{-0.00017}$ \\
$a/R_{\star}$     & $3.492^{+0.024}_{-0.025}$ & $3.562^{+0.022}_{-0.023}$ & $3.549^{+0.021}_{-0.021}$ \\
$i_{\rm{orb}}$    & $84.04^{+0.36}_{-0.38}$ & $84.88\pm0.33$ & $84.70^{+0.31}_{-0.30}$ \\
$u_{\rm{lin}}$     & $0.296\pm0.034$ & $0.2192^{*}$ & $0.2192^{*}$ \\
$u_{\rm{quad}}$ & $0.158^{+0.061}_{-0.060}$ & $0.3127^{*}$ & $0.3127^{*}$ \\
\hline
\multicolumn{4}{l}{$^{*}$ value taken from Claret (2017).} \\
}

In order to verify our procedure and its reliability of error estimates, the modelling was repeated for a scenario with a noncircular orbit with the initial conditions set as in Shporer \etal (2019). We entered $e_{\rm{orb}} = 0.0091$, $\omega = 269^{\circ}$ and the LD coefficients were fixed at the theoretical values $u_{\rm{lin}} = 0.2192$ and $u_{\rm{quad}} = 0.3127$ derived from Claret (2017). As it is shown in Table~2, our procedure reproduces the results of Shporer \etal (2019) very well. The parameters of our trial noncircular model agree with those of Shporer \etal (2019) within 0.2--0.4$\sigma$. The errors are similar for $i_{\rm{orb}}$ and $a/R_{\star}$, and our estimate of uncertainty for $R_p/R_{\star}$ appears to be greater by 30\%. As an additional test showed, this is a consequence of the inclusion of the quadratic term in the de-trending procedure.    

Differences between parameters of the circular-orbit model and those of Shporer \etal (2019) are noticeable at a 1.7--2.6$\sigma$ level. The values of both $i_{\rm{orb}}$ and $a/R_{\star}$ were found to be slightly smaller which is a direct consequence of the system's geometry. The transits were found to be deeper. The source of this effect is seen in the LD coefficients, which we set as the free parameters of the model. Because of inconsistencies of theoretical stellar limb darkening tables, it is advocated to keep the LD coefficients free in modelling of transit light curves if the photometry is of sufficient quality and these coefficients can be determined reliably (Csizmadia \etal 2013). We notice that inclusion of the uncertainties of the LD coefficients in the error budget increased errors of the other parameters, making them more reliable. Our model yields $u_{\rm{lin}}$ greater by 2.2$\sigma$ and $u_{\rm{quad}}$ smaller by 2.6$\sigma$ if compared to the theoretical expectations from Claret (2017). We note that the similar though less significant trend can be found for $V$-band data in Southworth \etal (2009).

%%%%%%%%%%%%%%%%%%%%%%%%%%%%%%%%%%%%%%%%%%%%%%%%%%%%%%%%%%%%%%%%%%%%%%

\section{Transit timing}\label{Sect.TT}

The transit model obtained in Sect.~3 was used as a template for the ground-based light curves in order to determine their mid-transit times. We used two follow-up light curves from Heller \etal (2009), five from Southworth \etal (2009), and two from Kedziora-Chudczer \etal (2019). Timestamps were converted to $\rm{BJD_{TDB}}$ and if needed magnitudes were rescaled into fluxes normalised to unity outside the transits. The photometric time series from Maxted \etal (2013), Wilkins \etal (2017), Cort\'es-Zuleta \etal (2020), and Patra \etal (2020) were not available. 

In addition, we acquired a new transit light curve on September 26, 2019 using the 0.6 m Helen Sawyer Hogg (HSH) telescope located at Complejo Astronomico El Leoncito (CASLEO, San Juan, Argentina). An SBIG STL-1001E CCD camera with $1024 \times 1024 \times 9~ \mu \rm{m}$ pixels was used as a detector. The instrument offered a field of view of $9.3 \times 9.3$ arc minutes. The light curve was acquired through an $I$-band filter with exposure times of 20--25 s (depending on seeing conditions), giving an average cadence of 32 s. The observations were reduced with a standard procedure carried out with AstroImageJ software (Collins \etal 2017). The fluxes were obtained with the aperture photometry method with an aperture radius, a set of comparison stars, and de-trending parameters being optimised to obtain the lowest noise. The final light curve together with a transit model is presented in Fig.~3.

% FIGURE 
\begin{figure}[thb]
\begin{center}
\includegraphics[width=0.7\textwidth]{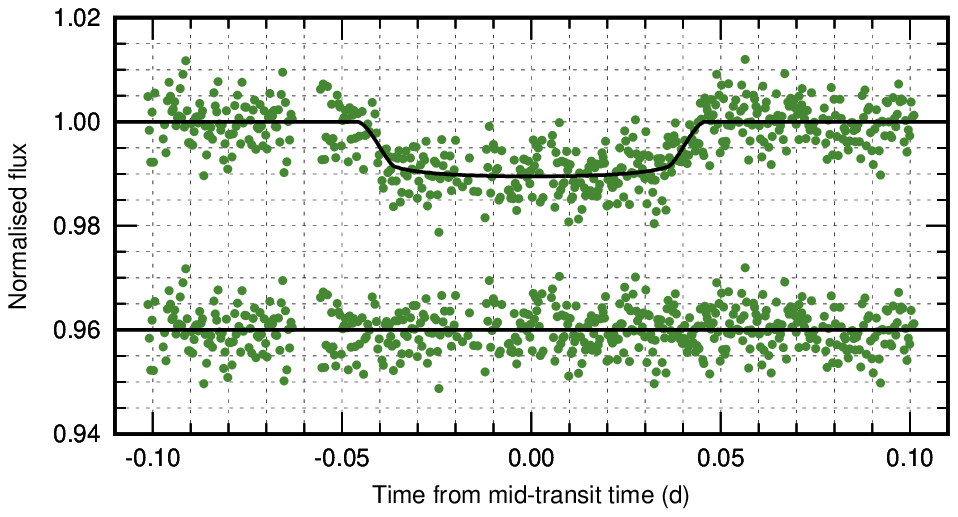}
\end{center}
\FigCap{Our new transit light curve observed on September 26, 2019 with the HSH telescope. Individual measurements are marked with dots, the best-fitting model is marked with a line. The quality of the light curve is degraded mainly by faintness of comparison stars available in the field of view. The photometric scatter is 3.0 parts per thousand of the normalised flux per minute of observation. The residuals are shown below.}
\end{figure}

The mid-transit times were determined with TAP. For each light curve, the TESS transit model with parameters obtained in Sect.~3 was fitted with the MCMC procedure using 10 chains of a length of $10^6$ steps. The transit parameters -- $R_p/R_{\star}$, $a/R_{\star}$, and $i_{\rm{orb}}$ -- were allowed to vary under Gaussian penalties of the template model. The LD coefficients were bi-linearly interpolated from tables of Claret \& Bloemen (2011) for stellar parameters determined by Heller \etal (2009), and also allowed to vary under a Gaussian prior of a width of 0.1. The coefficients of a second-order polynomial, which accounts for a possible trend in the time domain, and $T_{\rm{mid}}$ were kept free.

A signature of transits of WASP-18A~b was detected in broadband $Hipparcos$ photometry by McDonald \& Kerins (2018). The star was sparsely sampled between December 1989 and March 1993 with 130 measurements in total. Such early epochs are especially important for timing studies. The data were extracted from \textit{Hipparcos Epoch Photometry}, a complement to \textit{The Hipparcos and Tycho Catalogues} (ESA 1997), and then phase folded following a preliminary ephemeris. Phase of each data point was transformed into BJD$_{\rm{TDB}}$ of an artificial transit located near a middle of the time span of the observations. We note that the orbital period could be in principle a subject of variation but the scale of this variation is expected to be relatively small in a time scale of a few months or years, and any cumulative shift in transit times would be preserved. A mid-transit time, which is determined in this manner, is de facto an average representative for a time covered by the phase-folded observations. The magnitudes were transformed into fluxes and normalised using a median value of magnitude. The final light curve was trimmed to $\pm5$ hours around the expected mid-transit time. 

A similar procedure was applied to the All Sky Automated Survey (ASAS, Pojma\'nski 1997) and SuperWASP (Butters \etal 2010) photometry. The richest dataset extracted from the ASAS Photometric Catalog for an optimal aperture provides 525 measurements spread between November 2000 and December 2009 with a median cadence of 3 days. After trimming and applying an iterative 5$\sigma$ clipping of outlying data points, 230 measurements were qualified for further analysis. The SuperWASP database provides 3360 and 4330 observations done in 2006 and 2007, respectively. Because of the large number of data points and a high cadence, data in both observing seasons were analysed separately. Trimming and 5$\sigma$ clipping left the final light curves with 1430 and 2060 data points for 2006 and 2007, respectively. A subsequent procedure of the analysis was similar to that one which was applied to the single follow-up light curves. The only differences were that no time-domain trends were considered and the parameter $R_p/R_{\star}$ was allowed to float in order to prevent underestimation of errors. 

The compilation of mid-transit times is listed in Table~3. As the photometric time series from Maxted \etal (2013), Wilkins \etal (2017), Cort\'es-Zuleta \etal (2020), and Patra \etal (2020) were unavailable, the original mid-transit times were taken then. Southworth \etal (2010) note that the time stamps in photometric time series of Southworth \etal (2009) might be offset by an unknown amount because a clock of a computer which was used to record observations was not synchronised properly. We used the photometric data which are available via \textit{CDS} and converted their midpoints of observations given in BJD$_{\rm{UTC}}$ into BJD$_{\rm{TDB}}$. However, we found no systematic shift of this subset of mid-transit times. This finding indicates that the observations of Southworth \etal (2009) were practically unaffected by the faulty time service.

\MakeTable{ r l l l l}{12.5cm}{Mid-transit times for WASP-18 b.}
{\hline
 Epoch & $T_{\rm{mid}}$ (BJD$_{\rm{TDB}}$) & \multicolumn{1}{c}{$+\sigma$ (d)} & \multicolumn{1}{c}{$-\sigma$ (d)} & Light curve source\\
\hline
-6283 & 48466.366 & 0.033 & 0.021 & \textit{Hipparcos}, ESA (1997) \\
 -747 & 53518.2126 & 0.0084 & 0.0067 & ASAS, Pojma\'nski (1997) \\
 -267 & 53970.11352 & 0.00092 & 0.00100 & SuperWASP, Butters \etal (2010) \\
  156 & 54368.3462 & 0.0010 & 0.0011 & SuperWASP, Butters \etal (2010) \\
  471 & 54664.90568 & 0.00058 & 0.00058 & Heller \etal (2009) \\
  472 & 54665.84803 & 0.00063 & 0.00064 & Heller \etal (2009) \\
  915 & 55082.910597 & 0.00060 & 0.00063 & Southworth \etal (2009) \\
  916 & 55083.852439 & 0.00032 & 0.00030 & Southworth \etal (2009) \\
  917 & 55084.793919 & 0.00024 & 0.00024 & Southworth \etal (2009) \\
  918 & 55085.734997 & 0.00027 & 0.00026 & Southworth \etal (2009) \\
  919 & 55086.677153 & 0.00034 & 0.00035 & Southworth \etal (2009) \\
  969 & 55133.7472 & 0.0012$^{*}$ & 0.0012$^{*}$ & Cort\'es-Zuleta \etal (2020) \\
  970 & 55134.6914 & 0.0012$^{*}$ & 0.0012$^{*}$ & Cort\'es-Zuleta \etal (2020) \\
  971 & 55135.6331 & 0.0012$^{*}$ & 0.0012$^{*}$ & Cort\'es-Zuleta \etal (2020) \\
 1062 & 55221.30420$^{*}$ & 0.00010$^{*}$ & 0.00010$^{*}$ & Maxted \etal (2013) \\
 1286 & 55432.18970$^{*}$ & 0.00010$^{*}$ & 0.00010$^{*}$ & Maxted \etal (2013) \\
 1327 & 55470.78850$^{*}$ & 0.00040$^{*}$ & 0.00040$^{*}$ & Maxted \etal (2013) \\
 1330 & 55473.61440$^{*}$ & 0.00090$^{*}$ & 0.00090$^{*}$ & Maxted \etal (2013) \\
 1416 & 55554.57860$^{*}$ & 0.00050$^{*}$ & 0.00050$^{*}$ & Maxted \etal (2013) \\
 1433 & 55570.58400$^{*}$ & 0.00045$^{*}$ & 0.00048$^{*}$ & Maxted \etal (2013) \\
 1689 & 55811.5970 & 0.0041$^{*}$ & 0.0041$^{*}$ & Cort\'es-Zuleta \etal (2020) \\
 1758 & 55876.5559$^{*}$ & 0.0013$^{*}$ & 0.0013$^{*}$ & Maxted \etal (2013) \\
 2841 & 56896.14780$^{*}$ & 0.00080$^{*}$ & 0.00080$^{*}$ & Wilkins \etal (2017) \\
 3223 & 57255.78320$^{*}$ & 0.00030$^{*}$ & 0.00029$^{*}$ & Wilkins \etal (2017) \\
 3291 & 57319.80100$^{*}$ & 0.00039$^{*}$ & 0.00038$^{*}$ & Wilkins \etal (2017) \\
 3311 & 57338.6296 & 0.0011 & 0.0011 & Kedziora-Chudczer \etal (2019) \\
 3312 & 57339.57210 & 0.00052 & 0.00051 & Kedziora-Chudczer \etal (2019) \\
 3649 & 57656.84078 & 0.00097$^{*}$ & 0.00097$^{*}$ & Cort\'es-Zuleta \etal (2020) \\
 3650 & 57657.78359 & 0.00097$^{*}$ & 0.00097$^{*}$ & Cort\'es-Zuleta \etal (2020) \\
 3651 & 57658.72404 & 0.00097$^{*}$ & 0.00097$^{*}$ & Cort\'es-Zuleta \etal (2020) \\
 3684 & 57689.79147 & 0.00075$^{*}$ & 0.00075$^{*}$ & Patra \etal (2020) \\
 4042 & 58026.8319 & 0.0011$^{*}$ & 0.0011$^{*}$ & Cort\'es-Zuleta \etal (2020) \\
 4390 & 58354.45788 & 0.00019 & 0.00019 & TESS \\
 4391 & 58355.39931 & 0.00018 & 0.00018 & TESS \\
 4392 & 58356.34077 & 0.00022 & 0.00021 & TESS \\
 4393 & 58357.28206 & 0.00022 & 0.00024 & TESS \\
 4394 & 58358.22352 & 0.00021 & 0.00021 & TESS \\
 4395 & 58359.16514 & 0.00018 & 0.00017 & TESS \\
 4396 & 58360.10664 & 0.00019 & 0.00019 & TESS \\
 4397 & 58361.04799 & 0.00021 & 0.00022 & TESS \\
 4398 & 58361.98976 & 0.00021 & 0.00022 & TESS \\
 4399 & 58362.93133 & 0.00019 & 0.00019 & TESS \\
 4400 & 58363.87260 & 0.00019 & 0.00018 & TESS \\
 4401 & 58364.81379 & 0.00021 & 0.00020 & TESS \\
 4402 & 58365.75525 & 0.00019 & 0.00019 & TESS \\
\hline
\multicolumn{5}{l}{$^{*}$ Value taken from a source paper.}\\
}

\addtocounter{table}{-1}

\MakeTable{ r l l l l}{12.5cm}{Concluded.}
{\hline
 Epoch & $T_{\rm{mid}}$ (BJD$_{\rm{TDB}}$) & \multicolumn{1}{c}{$+\sigma$ (d)} & \multicolumn{1}{c}{$-\sigma$ (d)} & Light curve source\\
\hline
 4403 & 58366.69706 & 0.00021 & 0.00021 & TESS \\
 4406 & 58369.52127 & 0.00021 & 0.00021 & TESS \\
 4407 & 58370.46273 & 0.00020 & 0.00020 & TESS \\
 4408 & 58371.40404 & 0.00016 & 0.00016 & TESS \\
 4409 & 58372.34541 & 0.00020 & 0.00020 & TESS \\
 4410 & 58373.28729 & 0.00018 & 0.00018 & TESS \\
 4411 & 58374.22820 & 0.00017 & 0.00016 & TESS \\
 4412 & 58375.16982 & 0.00018 & 0.00017 & TESS \\
 4413 & 58376.11117 & 0.00021 & 0.00022 & TESS \\
 4414 & 58377.05268 & 0.00017 & 0.00017 & TESS \\
 4415 & 58377.99448 & 0.00019 & 0.00020 & TESS \\
 4416 & 58378.93581 & 0.00020 & 0.00019 & TESS \\
 4417 & 58379.87712 & 0.00020 & 0.00021 & TESS \\
 4418 & 58380.81887 & 0.00020 & 0.00020 & TESS \\
 4424 & 58386.46725 & 0.00019 & 0.00018 & TESS \\
 4425 & 58387.40877 & 0.00022 & 0.00023 & TESS \\
 4426 & 58388.35028 & 0.00020 & 0.00020 & TESS \\
 4427 & 58389.29173 & 0.00019 & 0.00019 & TESS \\
 4428 & 58390.23329 & 0.00017 & 0.00017 & TESS \\
 4429 & 58391.17446 & 0.00021 & 0.00021 & TESS \\
 4430 & 58392.11604 & 0.00019 & 0.00019 & TESS \\
 4431 & 58393.05742 & 0.00020 & 0.00021 & TESS \\
 4432 & 58393.99900 & 0.00019 & 0.00019 & TESS \\
 4433 & 58394.94026 & 0.00024 & 0.00023 & TESS \\
 4435 & 58396.82313 & 0.00020 & 0.00019 & TESS \\
 4436 & 58397.76454 & 0.00018 & 0.00018 & TESS \\
 4437 & 58398.70650 & 0.00017 & 0.00017 & TESS \\
 4438 & 58399.64746 & 0.00018 & 0.00018 & TESS \\
 4439 & 58400.58907 & 0.00018 & 0.00018 & TESS \\
 4440 & 58401.53084 & 0.00018 & 0.00019 & TESS \\
 4441 & 58402.47209 & 0.00019 & 0.00019 & TESS \\
 4442 & 58403.41357 & 0.00020 & 0.00020 & TESS \\
 4443 & 58404.35494 & 0.00019 & 0.00019 & TESS \\
 4444 & 58405.29602 & 0.00020 & 0.00020 & TESS \\
 4813 & 58752.69300 & 0.00089 & 0.00089 & this paper \\
\hline
\multicolumn{5}{l}{Table 2 in a machine-readable format is available at}\\
\multicolumn{5}{l}{http://www.home.umk.pl/\~gmac/TTV/doku.php?id=download or via CDS.}\\
}

The transit timing analysis was performed following a procedure adopted from Maciejewski \etal (2018). The MCMC algorithm running 100 chains, $10^4$ steps long each with the first 1000 trials discarded, was employed to refine the linear transit ephemerides 
\begin{equation}
     T_{\rm mid }(E)\,[{\rm BJD_{TDB}}] = 2454221.48183(8) + 0.94145242(2) \times E \, , \;
\end{equation}
where $E$ is a transit number counted from a reference epoch given in Hellier \etal (2009). The posterior probability distributions of the fitted parameters were used to determine their best-fitting values (medians) and their uncertainties (15.9 and 84.1 percentile values of the cumulative distributions). The best-fitting solution yields $\chi^2 = 66.9$ and ${\rm BIC_{lin}} = 75.6$.

A trial fit of a quadratic ephemeris in a form
\begin{equation}
  T_{\rm{mid}}= T_0 + P_{\rm{orb}} \times E + \frac{1}{2} \frac{d P_{\rm{orb}}}{d E} \times E^2 \, , \;
\end{equation}
where $T_0$ is the reference mid-transit time for the transit number 0 and $\frac{d P_{\rm{orb}}}{d E}$ is the change in the orbital period between succeeding transits, yields $\frac{d P_{\rm{orb}}}{d E} = (0.1 \pm 1.1) \times 10^{-10}$ days per epoch, $\chi^2 = 66.9$, and ${\rm BIC_{quad}} = 80.0$. The quadratic term is indistinguishable from zero well within 1$\sigma$ and the quadratic ephemeris is unambiguously disfavoured by the statistics. 

The timing residuals against the linear ephemeris are plotted in Fig.~4 together with uncertainties of the quadratic term.

% FIGURE 
\begin{figure}[thb]
\begin{center}
\includegraphics[width=1.0\textwidth]{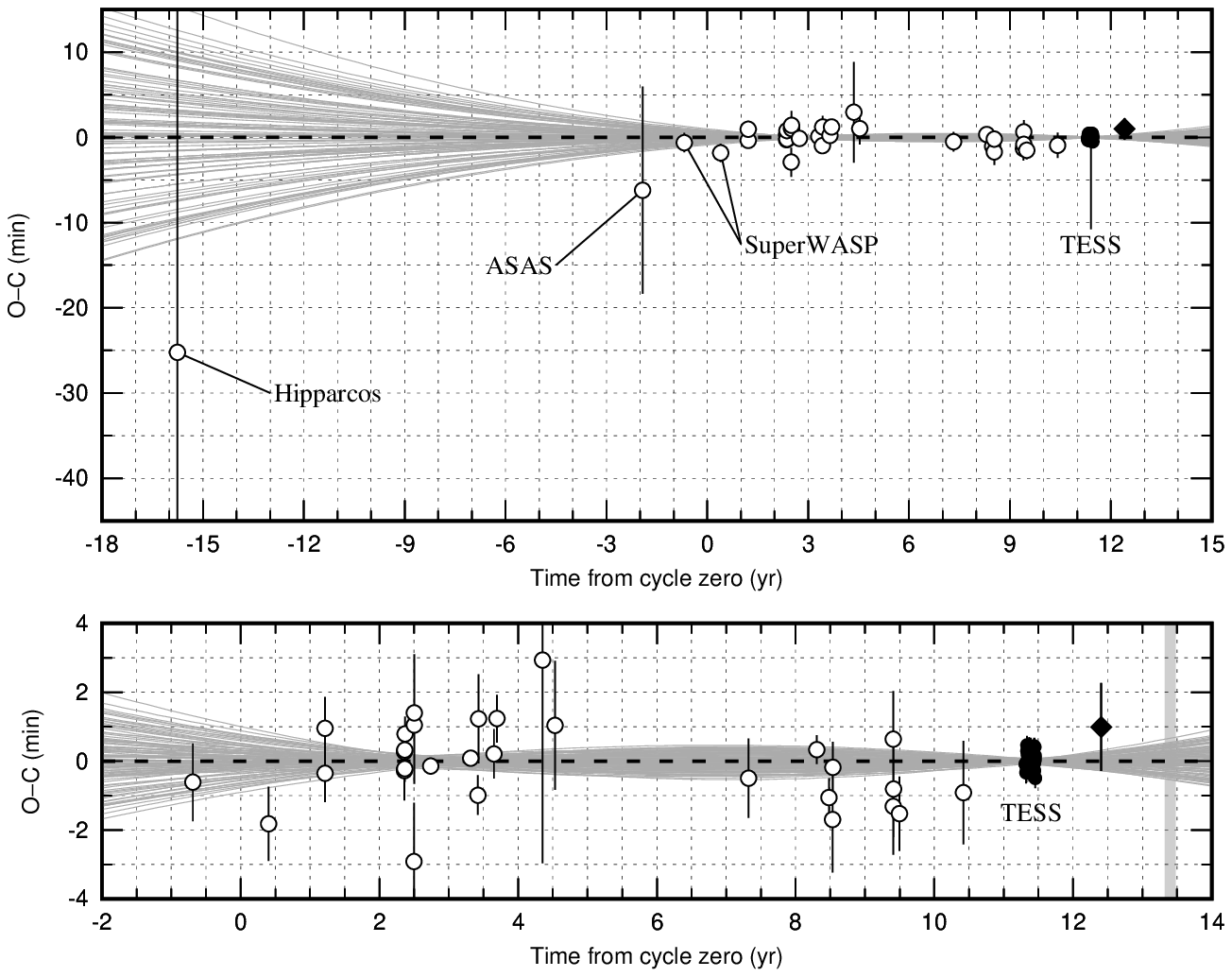}
\end{center}
\FigCap{Transit-timing residuals against the refined linear ephemeris. The new mid-transit time acquired with the HSH telescope is marked with a diamond. The re-determined mid-transit times from TESS are marked with dots. Data points from the literature are marked with open circles. The grey lines illustrate uncertainty of the trial quadratic ephemeris and are drawn for 100 sets of parameters, randomly chosen from the Markov chains. Lower panel shows the transit-timing residuals zoomed in on observations acquired in the last 2 decades. The greyed vertical strap marks additional TESS observations which are scheduled in August--November 2020.}
\end{figure}

The parameter $\frac{d P_{\rm{orb}}}{d E}$ is related to $Q'_{\star}$ with the formula (see \eg Maciejewski \etal 2018 and references therein)
\begin{equation}
  Q'_{\star} = - \frac{27}{2}\pi \left( \frac{M_{\rm p}}{M_{\star}}\right) {\left( \frac{a}{R_{\star}} \right)}^{-5} {\left( \frac{d P_{\rm{orb}}}{d E} \right)}^{-1} P_{\rm{orb}} \, , \;
\end{equation}
where $M_{\rm p}$ is a planet mass and $M_{\star}$ is a mass of a host star mass. Since no orbital decay was detected for WASP-18A~b, the lower constraint on $Q'_{\star}$ at the 95\% confidence level can be placed from the 5th percentile of the posterior probability distribution of $\frac{d P_{\rm{orb}}}{d N_{\rm tr}}$. Adopting the stellar mass $M_{\star} = 1.25 \pm 0.13$ (Hellier \etal 2009) and taking the remaining quantities determined in this study, we obtained that $Q'_{\star} > 3.9 \times 10^6$.

%%%%%%%%%%%%%%%%%%%%%%%%%%%%%%%%%%%%%%%%%%%%%%%%%%%%%%%%%%%%%%%%%%%%%%

\section{Discussion}

In all studies addressing the issue of the orbital eccentricity of WASP-18A~b, its value was found to be non-zero within 2.5-8.9$\sigma$, depending on the data used and the methodology of their analysis (Hellier \etal 2009, Triaud \etal 2010, Nymeyer \etal 2011, Knutson \etal 2014). Our analysis yields $e_{\rm b}$ which differs from 0 at a 10$\sigma$ level. In Maciejewski \etal (2020), we show that our procedure of the RV analysis provides reliable, not underestimated uncertainties. Following Eq.\ (3) of Adams \& Laughlin (2006), a circularisation timescale for WASP-18A~b is $\sim$20 Myr using a conservative value of the planetary quality factor of $10^6$. This is more than one order of magnitude shorter than the age of the system, which falls in a range 0.5--1.5 Gyr (Hellier \etal 2009). Thus, the non-zero eccentricity is rather unexpected unless there is an efficient mechanism which excites and maintains it or the value of the planetary quality factor is underestimated. The apparently non-zero eccentricity together with the improbable orientation of the line of apsides, which appear to be aligned nearly exactly with the line of sight, can be naturally explained with the tidal RV signature.

Employing the equilibrium tide approximation, Arras \etal (2012) predict the tidal RV amplitude to be $\sim$32~m~s$^{-1}$. Using their Eq.\ (20), which arises from a simplification of general considerations for a circular orbit, and adopting the stellar mass $M_{\star} = 1.25 \pm 0.13$ (Hellier \etal 2009) and taking the remaining quantities determined in this study we found, however, that the predicted tidal RV amplitude is $\sim$20~m~s$^{-1}$. The main source of uncertainties in the theoretical predictions is a parameter $f_2$, which contains information on LD, and is proportionally related to $K_{\rm{tide}}$. In our calculations, we followed Arras \etal (2012) who used the LD under the Eddington approximation, for which $f_2 \approx 1.1$. Since the equilibrium tide approximation is supposed to be accurate to a factor of about 2 (Arras \etal 2012), the empirically determined value of $K_{\rm{tide}}=17.9\pm1.9$~m~s$^{-1}$ can be considered as being consistent with the theoretical prediction. After transformation of Eq.\ (20) of Arras \etal (2012), we obtained 
\begin{equation}
  f_2 = \frac{1}{3 \pi} \frac{K_{\rm tide}}{R_{\star}} \frac{M_{\star}}{M_{\rm b}} \left( \frac{a}{R_{\star}} \right)^{3} P_{\rm orb} \left( \sin i_{\rm orb} \right)^{-2} = 0.98 \pm 0.12\,  \;
\end{equation}
for WASP-18A. This result agrees with the Eddington approximation at the 1$\sigma$ level.
 
A rough estimate of a hight of the tides relative to the unperturbed stellar radius comes from the ratio of the tidal acceleration to the star's surface gravity (Pfahl \etal 2008)
\begin{equation}
  H_{\rm exp} = \frac{M_{\rm b}}{M_{\star}} \left( \frac{a}{R_{\star}} \right)^{-3} R_{\star} \, . \;
\end{equation}
For the WASP-18 system, we obtained $H_{\rm exp} = 159 \pm 10 {\rm ~km}$. This quantity can be empirically determined by calculating a distance traveled by a stellar photosphere in a time $t$ equal to a quarter of a tidal cycle  
\begin{equation}
  H_{\rm tide} = p K_{\rm{tide}} \int_{0}^{\frac{1}{4}P_{\rm tide}} \sin \left( \frac{2\pi}{P_{\rm tide}} t \right) {\rm{d}} t\, , \;
\end{equation}
where $p = 1.36$ is a projection factor, which scales disk integrated RVs into actual photospheric velocities (Getting 1934, Burki \etal 1982), and $P_{\rm tide} = P_{\rm orb} / 2$. We obtained $H_{\rm tide} = 157 \pm 16$~km which is consistent with $H_{\rm exp}$ well within 1$\sigma$.

Using Eq. (8) of Shporer \etal (2019), we redetermined the expected amplitude of the photometric modulation $A_{\rm ellip} = 199 \pm 12$ ppm. On the other hand, simple geometrical considerations lead to the relation
\begin{equation}
  A_{\rm ellip} = \frac{H_{\rm tide}}{R_{\star}-H_{\rm tide}}  \, , \;
\end{equation}
which yields $A_{\rm ellip} = 181 \pm 21$ ppm. This value, which was de facto derived from the RV tides, is in excellent agreement with both the theoretical predictions and the observed amplitude of $190.0^{+5.9}_{-5.8}$ ppm (Shporer \etal 2019). 
 
Csizmadia \etal (2019) has recently postulated that the eccentric orbit of WASP-18A~b is undergoing apsidal precession. The rate of this precession was found to be $\dot{\omega}=0.0091^{+0.0040}_{-0.0018}$ degrees per day using the literature RV together with the transit and occultation timing datasets. While trying to reproduce this result, we noticed that the best-fitting solution is found for $\dot{\omega}=0.0033 \pm 0.0071$ degrees per day. No local minimum of a $\chi^2$ distribution was found around the value of $\dot{\omega}$ postulated by Csizmadia \etal (2019). Compared to the model with no precession (Sect.~2), the precession is disfavoured by the statistics. The non-precession model gives $\chi^2 = 490.0$ and ${\rm BIC} = 533.0$ with no jitter added. Although the precession model yields a slightly smaller $\chi^2 = 485.1$, engaging $\dot{\omega}$ as the additional free parameter results in no significant improvement in ${\rm BIC} = 532.8$. Adding jitter results in a decrease of $\dot{\omega}$ down to $\dot{\omega}=-0.0015 \pm 0.0016$ for the jitter of 10.7~m~s$^{-1}$.  
A negative value of $\dot{\omega}$ cannot be induced by the tidal deformations. It could be produced in a result of the rotational deformation of the rapidly rotating host star if the planetary orbit were significantly misaligned (Migaszewski 2012). However, the rotation period of WASP-18A of $\approx 5.5$~d (Csizmadia \etal 2019) is significantly longer than the value of about 7 hours that would be required to reproduce $\dot{\omega} \approx -0.0015$ with WASP-18A~b on a polar orbit. Furthermore, the orbit of the planet was found to be well aligned (Triaud \etal 2010, Albrecht \etal 2012). As the redetermined value of $\dot{\omega}$ is consistent with zero regardless the amount of jitter added, the detection of apsidal precession of WASP-18A~b appears to be premature.

As noted by McDonald \& Kerins (2018), early \textit{Hipparcos} observations of WASP-18A~b provide rather weak constraints in transit timing studies. In a test run with those data skipped, the constraint on $Q'_{\star}$ was found to differ by a marginal value of 3\%. While our value of the timing residual of about $-0.018$ d is consistent with $-0.021$ d derived from the mid-transit time reported by McDonald \& Kerins (2018), our timing errors were found to be 2.5--2.6 times greater. To check if our procedure overestimates timing uncertainties, we calculated a ratio of our timing errors for TESS data to errors derived by Shporer \etal (2019). We found that this ratio is between 0.93 and 1.36 with a median value of 1.19. Thus, we conclude that the timing errors reported by McDonald \& Kerins (2018) might be underestimated. We also note that the timing errors from the \textit{Hipparcos} photometry are significantly asymmetric with $\sigma^{+}/\sigma^{-} = 1.6$. This effect is also visible in the original results of McDonald \& Kerins (2018) with $\sigma^{+}/\sigma^{-} = 1.5$. This asymmetry is a consequence of a non-uniform data point distribution in the \textit{Hipparcos} light curve.     

Our homogenous transit-timing analysis has provided the tightest constraint on $Q'_{\star}$ for WASP-18A. Wilkins \etal (2017) used all timing data available then and got $Q'_{\star}>10^6$ at 95\% confidence. Although McDonald \& Kerins (2018) obtained $Q'_{\star} \approx 5 \times 10^5$, Shporer \etal (2019) could only place a constraint on $Q'_{\star}> 1.7 \times 10^6$. The same constraint has been obtained recently by Patra \etal (2020). Our determination of $Q'_{\star}$ excludes values smaller than $3.9 \times 10^6$ with 95\% confidence. However, it is still not enough to verify theoretical findings of Collier Cameron \& Jardine (2018). They predict that the stars hosting hot Jupiters could have $Q'_{\star}$ of the order of $2 \times 10^8$ if the equilibrium-tide regime is considered. Under favourable circumstances, the dynamical-tide mechanism could operate in stellar interiors and the efficiency of tidal dissipation would be boosted by one order of magnitude that translates into $Q'_{\star} \approx 2 \times 10^7$.

WASP-18A will be observed again with TESS between August and November 2020. As it is shown in Fig.~4 (lower panel), new mid-transit times will definitely place tighter constraint on $Q'_{\star}$. Adopting $Q'_{\star} = 2 \times 10^7$, a departure of 1 minute from a linear ephemeris could be detected after 2 decades of precise observations. However, it is more likely that the system is far from the dynamical-tide regime and the host star dissipates the tidal energy less efficiently. In such case, the cumulative time shift of 1 minute would be noticed after about 60 years. Nevertheless, the WASP-18 system still remains one of the best candidates for an infalling hot Jupiter orbiting a main sequence star. The rapid decay rate of the WASP-12~b, so far the only planet for which the orbital evolution due to tidal interactions has been observed (Maciejewski \etal 2016, 2018, Maciejewski 2019, Yee \etal 2020), seems to be triggered by the evolutionary changes in the star's interior structure (Weinberg \etal 2017).

%%%%%%%%%%%%%%%%%%%%%%%%%%%%%%%%%%%%%%%%%%%%%%%%%%%%%%%%%%%%%%%%%%%%%%

\section{Conclusions}

As with the WASP-12 system (Maciejewski \etal 2020), the observed variation in RVs of WASP-18A can be decomposed into the component induced by the orbital motion of the planet and the signal produced by the motion of the photosphere following the planetary tidal potential. The amplitude of these RV tides was found to agree with both the predictions of the equilibrium tide approximation and the ellipsoidal modulation observed in the space-borne orbital phase curve. The orbit of the planet appears to be de facto circular making the apsidal precession beyond possibility of detection.

Although the WASP-18 system is one of the top candidates for which loss of orbital angular momentum due to dissipation of planetary tides in the host star could be observed, planetary in-spiralling remains undetected. Transit timing data indicate that the modified tidal quality parameter $Q'_{\star}$ of WASP-18A must be greater than the canonical value of $10^6$ reported in studies of binary stars in stellar clusters (Meibom \& Mathieu 2005, Ogilvie \& Lin 2007, Milliman \etal 2014). This finding is in line with recent studies on the population of hot Jupiters (Collier Cameron \& Jardine 2018). Further systematic timing observations acquired within a decade are expected to permit for probing values of $Q'_{\star}$ from the dynamical-tide regime.

%%%%%%%%%%%%%%%%%%%%%%%%%%%%%%%%%%%%%%%%%%%%%%%%%%%%%%%%%%%%%%%%%%%%%%

\Acknow{We are grateful to Dr.~Coel Hellier and Dr.~Lucyna Kedziora-Chudczer for sharing the light curves with us. GM and CM acknowledge the financial support from the National Science Centre, Poland through grant no. 2016/23/B/ST9/00579. This work was based on observations at the W.\ M.\ Keck Observatory granted by the University of Hawaii, the University of California, and the California Institute of Technology. We thank the observers who contributed to the measurements reported here and acknowledge the efforts of the Keck Observatory staff. We extend special thanks to those of Hawaiian ancestry on whose sacred mountain of Mauna Kea we are privileged to be guests. This paper includes data collected with the TESS mission, obtained from the MAST data archive at the Space Telescope Science Institute (STScI). Funding for the TESS mission is provided by the NASA Explorer Program. STScI is operated by the Association of Universities for Research in Astronomy, Inc., under NASA contract NAS 5-26555. This paper makes use of data from the DR1 of the WASP data (Butters \etal 2010) as provided by the WASP consortium, and the computing and storage facilities at the CERIT Scientific Cloud, reg. no. CZ.1.05/3.2.00/08.0144 which is operated by Masaryk University, Czech Republic.}

%%%%%%%%%%%%%%%%%%%%%%%%%%%%%%%%%%%%%%%%%%%%%%%%%%%%%%%%%%%%%%%%%%%%%%

\end{document}